\documentclass[aps,floats,pre]{revtex4}
\usepackage{graphicx}
\usepackage{amssymb}
\begin{document}
\title{Front Propagation Dynamics with Exponentially-Distributed Hopping}

\author{Elisheva Cohen} 
\affiliation{Dept. of Physics, Bar-Ilan  University, Ramat-Gan, IL52900 Israel}
\author{David A. Kessler}
\affiliation{Dept. of Physics, Bar-Ilan University, Ramat-Gan, IL52900 Israel}
\date{\today}
\begin{abstract}
  We study reaction-diffusion systems where diffusion is by
  jumps whose sizes are distributed exponentially.  We first
  study the Fisher-like problem of propagation of a front into an unstable
  state, as typified by the A+B $\to$ 2A reaction.  We find that
  the effect of fluctuations is especially pronounced at small hopping rates.
  Fluctuations are treated heuristically via a density cutoff in the
  reaction rate.
  We then consider the case of propagating up a
  reaction rate gradient. 
  The effect of fluctuations here is pronounced, with the front
  velocity increasing without limit with increasing bulk particle
  density.  The rate of increase is faster than in the case of a 
  reaction-gradient with nearest-neighbor hopping.  
  We derive analytic expressions for the front velocity dependence on
  bulk particle density. Compute simulations are performed to confirm the
   analytical results.
\end{abstract}
\maketitle

Many physical, chemical, and biological systems exhibit fronts which
propagate through space. Familiar examples range from chemical
reaction dynamics such as flames~\cite{kpp}, phase transitions
such as solidification~\cite{kkl}, the spatial spread of
infections~\cite{blumen}, and even the fixation of a beneficial allele
in a population~\cite{fisher}. It is thus
of great interest to understand the universality classes of fronts
which govern what will happen when systems such as these are prepared
in a spatially heterogeneous manner. These classes determine the
selection of propagation speed, the sensitivity to 
particle-number fluctuations, and the stability of the front with respect to
deviations from planarity.

The simplest kind of such a front
is that wherein a stable phase replaces a metastable
one~\cite{kkl}. Here the mean-field front velocity is determined via
the requirement that there exists a heteroclinic trajectory of the 
moving-frame
steady-state problem (wherein the solution depends only on $x-vt$) connecting the
 metastable
phase at $+\infty$ with the stable one at $-\infty$.  This type of
front is robust with respect to fluctuations, with power-law
corrections in $1/N$ (where $N$ is the number of particles per site in
the final state) to the mean-field limit~\cite{kns}.  A second class
is exemplified by the simple infection model $A+B \rightarrow 2A$ on a
1d lattice (with spacing $a$) with equal $A$ and $B$ hopping rates~\cite{blumen}; 
this process leads in
the mean-field limit to a spatially discrete version of the
 Fisher equation~\cite{fisher} 
\begin{equation}
\dot{\phi}(x) \ = r \phi(x) \left(1 - \phi(x)\right) + \frac{D}{a^2} \left(\phi(x+a) - 2\phi(x)
+ \phi(x-a)\right) . 
\label{fisher}
\end{equation}
Here propagation is into
the linearly unstable $\phi=0$ state, where $\phi$ is the number of
$A$ particles at a site. Recent
work~\cite{bd,kns,vansaarloos,pechenik} has shown that the front behavior in
the stochastic model does approach that of the Fisher equation, where
the velocity is selected by the (linear) marginal stability criterion~\cite{ben-jacob} to be $2\sqrt{rD}$, albeit with an
anomalously long transient $O(1/t)$ and anomalously large fluctuation
corrections $O(1/\ln ^2 N)$  . There are also some findings in
regard to both front stability in the case of unequal $D$~\cite{nature}, 
and also the scaling properties of front
fluctuations~\cite{moro}. Finally, there are also fronts 
which have properties intermediate to the previous two classes.

In a recent work~\cite{kess,kess1}, we introduced a new class of 
fronts corresponding to
propagation into an unstable state 
up a reaction-rate gradient~\cite{shapiro,freidlin}.  
This type of gradient is present, for example, in systems with
an inherent spatial 
inhomogeneity,
 and also in models
 of Darwinian evolution~\cite{tsimring,evol1,rouzine,sex}, (where the
birth rate, which is parallel to our reaction rate, is proportional to
 fitness $x$). 
We found that the sensitivity to fluctuations in the presence of such
a positive reaction-rate gradient is greatly enhanced.  In particular,
the front velocity diverges with increasing bulk particle density.  As a
corollary,
the standard reaction-diffusion equation treatment is not useful, as it
gives rise to finite-time singularities in the velocity.  Also, the velocity
is strongly sensitive to details of diffusion, with the increase
of the velocity with density being qualitatively stronger for a lattice
system than in the continuum.

Given this sensitivity to the precise implementation
of diffusion, in this work we turn to the study the effect of 
implementing diffusion via infinite-range hopping, where the
size of the jumps is distributed exponentially.  Such
a model has been considered, for example, in the description of the
airborne dispersion of seeds, leading to the spread of a particular
colony of plants. It is also relevant in the evolution context, where the change
in fitness  due to mutations is commonly assumed to be exponentially distributed~\cite{lenski}.
We will see that
even in the absence of a gradient, this form of diffusion increases 
dramatically the effect of fluctuations, at least for small hopping rates.
In particular, the naive reaction-diffusion formalism predicts a finite 
velocity in the limit of zero hopping rate, which is clearly unphysical. 
Introducing a reaction-rate gradient again changes the functional dependence
of the velocity on density from that of the nearest-neighbor hopping studied
previously. 

The plan of the paper is as follows.  In Section \ref{fisher_sec}, we discuss
the gradient-free model, and derive the velocity in the limit of infinite
density.  We show that for fixed hopping rate, the finite density correction
formula derived by Brunet and Derrida for the Fisher equation is applicable.
However, this formula breaks down in the small hopping rate limit.  We derive
an analytical expression for the velocity in this limit.  In Section
\ref{snyder_sec}, we discuss a similar model of Snyder designed to model the
spread of colonies, showing that the same physics applies upon the correct
mapping of parameters.  In Section \ref{grad_sec}, we introduce our reaction-rate
gradient model, and after briefly reviewing what is known for continuum diffusion and
nearest-neighbor hopping, we calculate an
analytical approximation to the velocity for large density. Finally, in 
Section \ref{final_sec}, we summarize our results and draw some conclusions.

\section{\label{fisher_sec}Exponential Hopping Fisher Equation}

In this model, the hopping probability has an unbounded range, and decreases exponentially with distance.
 As in the Fisher model~\cite{fisher}, the reaction rate is a constant $r$.
In the continuum limit, the equation describing this model is:
\begin{equation}
\label{inteqn}
\dot{\phi}(x)=\frac{D\beta^3}{2} \int_0^\infty ds e^{-\beta s} \left(\phi(x+s) + \phi(x-s)\right) - D\beta^2\phi(x) + r\phi(x)\left(1-\phi(x)\right),
\label{rd}
\end{equation}
and the steady-state equation is:
\begin{equation}
\label{inteqn1}
\frac{D\beta^3}{2} \int_0^\infty ds e^{-\beta s} \left(\phi(x+s) + \phi(x-s)\right) - D\beta^2\phi(x) + v\phi'(x) + r\phi(x)\left(1-\phi(x))
\right)= 0.
\end{equation}
It is useful to convert this equation into a differential equation using the fact that
\begin{equation}
O_\pm \int_0^\infty ds e^{-\beta s} \phi(x\pm s) \equiv \left(\beta\mp \frac{d}{dx}\right) \int_0^\infty ds e^{-\beta s} \phi(x\pm s) = \phi(x).
\end{equation}
Then, acting upon Eq. (\ref{inteqn1}) by $O_+ O_-$ yields
\begin{equation}
D\beta^2 \phi'' +  \left(\beta^2-\frac{d^2}{dx^2}\right)(v\phi' + r\phi(1-\phi) ) = 0
\label{basiceq}
\end{equation}
As with the standard Fisher equation, this equation has solutions for all velocities, and positive definite
solutions for all velocities greater than some critical velocity, the so-called marginally stable velocity, $v_F$,
which is the asymptotic velocity of propagation of all fronts with initial compact support.
This can be found from the dispersion relation for the leading edge where $\phi\sim e^{-kx}$:
\begin{equation}
D\beta^2 k^2 + (\beta^2-k^2)(-vk+r) = 0
\label{dispers}
\end{equation}
The marginally stable velocity, $v_F$, is then given by the requirement that Eq. (\ref{dispers})
has a degenerate solution, leading to the discriminant condition
\begin{equation}
0=\frac{d}{dk}\left[D\beta^2 k^2 + \left(\beta^2-k^2\right)(-vk+r)\right]=2D\beta^2 k - \beta^2 v + 3vk^2 - 2rk .
\label{discrim}
\end{equation}
Solving simultaneously Eqs. (\ref{dispers}) and (\ref{discrim})
 yields, introducing $t\equiv \sqrt{D^2\beta^4 + 8D\beta^2 r}$
\begin{equation}
v_F=\frac{\sqrt{2}(5D\beta^2+4r+3t)}{8\beta}\sqrt{\frac{D\beta^2+2r-t}{r-D\beta^2}}
\label{v0fish_exp}
\end{equation}
This has the scaling form $v_F=2\sqrt{rD}f(\beta^2 D/r)$ where the function $f(x)\to 1$ for $x \to \infty$ and
$f(x)\sim 1/(2\sqrt{x})$ as $x \to 0$.  Thus, for large $\beta$, we recover the usual Fisher answer.  What is remarkable
is that the velocity has the finite limit $r/\beta$ as $D \to 0$, so that we have velocity without diffusion!

\subsection{Calculation of the Velocity for a Small Cutoff}
This anomaly is yet another example of how the reaction diffusion equation, Eq. (\ref{rd}), provides incorrect
information about the original stochastic model.  A more accurate picture is achieved by studying a cutoff version
of the equation, wherein the reaction is turned off wherever $\phi(x)$ is less than some threshold $\epsilon$, of
order $1/N$\cite{mf-dla,kepler,tsimring,bd,kns}.  This captures an essential feature of the original model, namely that the reaction
zone always has compact support. Brunet and Derrida \cite{bd} have provided a general formula for the correction induced in 
the velocity due to the cutoff (for small cutoffs) for Fisher-like equations.  This formula reads
\begin{equation}
v_{\epsilon}=v_F-\frac{v''(k_F)\pi^2 k_F^2}{2(\ln\epsilon)^2}.
\label{bdfisher}
\end{equation}
where $k_F$ is the degenerate solution of the dispersion relation, Eq. (\ref{dispers}).  Although derived for
second order equations, whereas our equation is of third order, nevertheless, as we shall see, it correctly
gives the leading order correction for the velocity in our case as well.  

\subsubsection{Jump Conditions at the Cutoff Point}
The first task, as for the standard Fisher equation, is to solve the equation for the region beyond the cutoff, 
where $\phi(x) < \epsilon$.  This
will give a set of boundary conditions at the cutoff point, $x_c$.  Due to the third-order nature
of our equations, and that the derivatives act on the now discontinuous reaction term, these conditions
are fairly messly. 
While  the solution is continuous at the cutoff point,
 there is no continuity of the 
first and second derivatives at this point.  
 
 To derive the correct jump conditions, we start from the integral equation, Eq. (\ref{inteqn1}).
 For $x>x_c$, the solution is $\phi_x=\epsilon e^{-k_r(x-x_c)}$, where $k_r$ satisfies the $r=0$ dispersion relation
\begin{equation}
D\beta^2 k_r^2 - (\beta^2-k_r^2)vk_r = 0
\label{kreqn}
\end{equation}
or, 
\begin{equation}
k_r = \frac{-D\beta^2 + \sqrt{D^2\beta^4 + 4\beta^2 v^2}}{2v}
\end{equation}
Thus, 
\begin{equation}
\int_0^\infty ds e^{-\beta s} \phi(x_c + s) = \frac{\epsilon}{\beta+k_r}
\end{equation}
Evaluating the integral equation, Eq. (\ref{inteqn1}), as $x \to x_c^+$ gives
\begin{equation}
\frac{D\beta^3}{2}\left[ \frac{\epsilon}{\beta+k_r} + \int_0^\infty ds e^{-\beta s} \phi(x_c-s) \right] -D\beta^2\epsilon
 - vk_r\epsilon =0
\end{equation}
This, together with Eq. (\ref{kreqn}), yields
\begin{equation}
\int_0^\infty ds e^{-\beta s} \phi(x_c-s) = \frac{\epsilon}{\beta-k_r}
\end{equation}
Now, let us analyze Eq. (\ref{inteqn1}) for $x \to x_c^-$.  We get
\begin{equation}
\phi'(x_c^-) = -\frac{r\epsilon(1-\epsilon)}{v} - k_r\epsilon
\label{phitag}
\end{equation}
Breaking up Eq. (\ref{inteqn1}) as follows,
\begin{eqnarray}
0&=&\frac{D\beta^3}{2}\left[\int_0^{x_c-x} ds e^{-\beta s} \phi(x+s) + \int_{x_c-x}^\infty ds e^{-\beta s}\phi(x+s)
+ \int_0^\infty ds e^{-\beta s} \phi(x-s)\right]\nonumber\\
& -&D\beta^2\phi + v\phi' + r\phi(1-\phi) \nonumber\\
&=& \frac{D\beta^3}{2}\left[2\int_0^{x_c-x}ds \sinh(\beta s)\phi(x+s) + e^{-\beta(x_c-x)}\frac{\epsilon}{\beta+k_r}
+ e^{\beta(x_c-x)}\frac{\epsilon}{\beta-k_r}\right]\nonumber\\
& -& D\beta^2\phi + v\phi' + r\phi(1-\phi)
\end{eqnarray}
and taking a derivative, we get
\begin{eqnarray}
\frac{D\beta^3}{2} \left[ \beta\frac{\epsilon}{\beta+k_r}  - \beta\frac{\epsilon}{\beta-k_r}\right]
 -   D\beta^2\phi'(x_c^-)
+v \phi''(x_c^-) + r\phi'(x_c^-)(1-2\phi(x_c^-)) = 0
\end{eqnarray}
or
\begin{equation}
\phi''(x_c^-) =\epsilon \frac{-\beta^2 v r + v^2 k_r^3 + 2vrk_r^2 + r^2 k_r}{k_r v^2}
+ \epsilon^2\frac{\beta^2 v r - 3 vrk_r^2 - 3r^2 k_r}{k_r v^2} + \epsilon^3\frac{2r^2}{v^2}.  
\end{equation}

\subsubsection{The Modified BD Treatment}
As in the original BD treatment, we divide the range of $x<x_c$ into two regions. In the first region, $\phi(x)$ is not small compared to $1$, but the effect of the cutoff is negligible.
In the second region $\epsilon<\phi(x)<<1$. We fix the translation invariance by requiring $\phi(0)=1/2$. Then
as $\epsilon \to 0$, $x_c \to \infty$.  
In the first region, we can take the velocity to be $v_F$, so that there is a degenerate solution of the
dispersion relation.  Then, for large $x$, the dominant solution is

\begin{equation}
\phi(x)\sim Axe^{-k_F x} 
\label{1regsol}
\end{equation} 

In the second region, since the velocity is close to $v_F$, $v_\epsilon=v_F - \Delta$, $\Delta \ll 1$, the general solution is: 
\begin{equation}
\phi(x)\sim Be^{-k_F x} \sin(k_ix+C)+Fe^{-k_2x}.
\label{2regsol}
\end{equation}
where $k_2<0$ is the third (nondegenerate) root of the dispersion relation and
$k_i \sim \sqrt{\Delta}$ and we can ignore the $0(\Delta)$ shift in the real part of $k$.
Matching between the first 
and the second region requires that $B=A/k_i$, $C=0$ and $Fe^{-(k_2-k_F) x_c}\ll 1$.  Now, in general,
we have to enforce three jump conditions, (whose left hand sides are $\Delta$-independent to leading order),
 with the two free parameters $B$ and $F$, which is impossible.
The only way to make things work is to have $\sin(k_i x_c)$ be of the same order as $k_i \cos(k_i x_c)$,
in other words $k_i x_c \approx \pi - O(\Delta^{1/2})$, which is exactly the same condition as in the original
BD treatment, where there was one free parameter and two jump conditions.
Since
\begin{equation}
k_i = \sqrt{\frac{2\Delta}{v''(k_F)}}
\end{equation}
we immediately recover the BD result quoted above, Eq. (\ref{bdfisher}).
  
\begin{figure}  
\includegraphics[width=.4\textwidth]{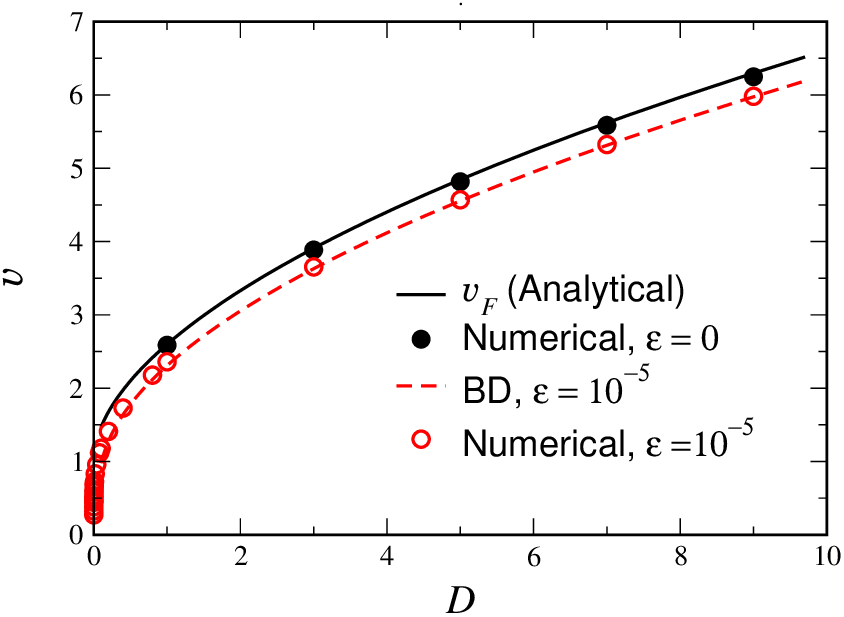} 
\caption{$v$ vs. large $D$ for exponential Fisher model. Analytical formula for $v_0$ (\ref{v0fish_exp}),
and analytical formula for $v_{\epsilon}$ (\ref{bdfisher}), compared to numerical results. $\epsilon=10^{-5}$, $\beta=1$. (color online)}
\label{fig1}
\end{figure}

\begin{figure}  
\includegraphics[width=.4\textwidth]{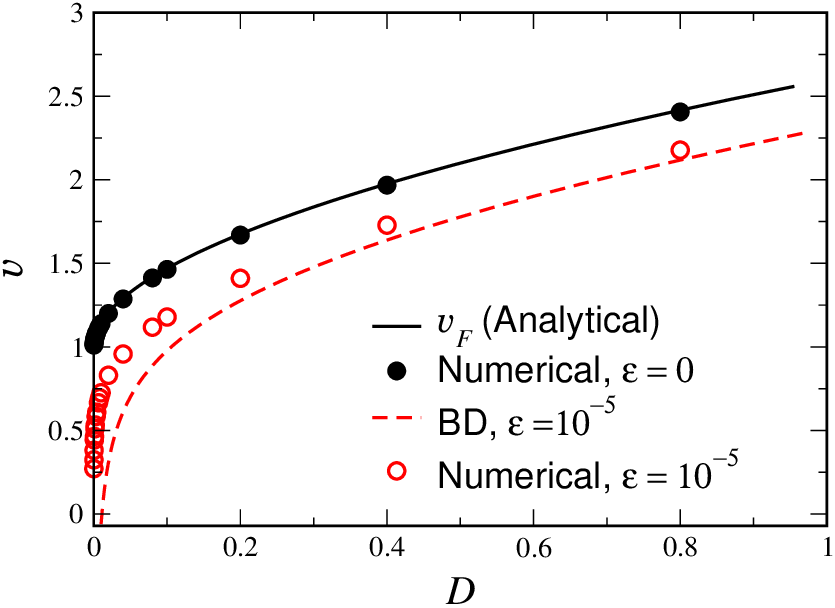} 
\caption{$v$ vs. small $D$ for exponential Fisher model. Analytical formula for $v_0$ (\ref{v0fish_exp}),
and analytical formula for $v_{\epsilon}$ (\ref{bdfisher}), compared to numerical results. $\epsilon=10^{-5}$, $\beta=1$ (color online)}
\label{fig2}
\end{figure}

Examining the BD result, we see that in the limit of  $D\beta^2/r \gg 1$, $k_F \sim \sqrt{r/D}$ and
$v''(k_F) \sim 2\sqrt{D^3/r}$, so that
\begin{equation}
v_\epsilon \sim v_F -\frac{\pi^2\sqrt{rD}}{\ln^2 \epsilon}
\end{equation}
which is of course the Fisher result.  On the other hand, when $D\beta^2/r \ll 1$, $k_F \sim \beta - \beta^2\sqrt{D/2r}$
and $v''(k_F) \sim (2r)^{3/2}/\beta^4/\sqrt{D}$, and so
\begin{equation}
v_\epsilon \sim v_F - \frac{\pi^2 (2r)^{3/2}}{2\beta^2\sqrt{D}\ln^2 \epsilon}
\end{equation}
Thus the BD correction diverges as $D\to 0$.  Thus, while for sufficiently small $\epsilon$, the BD correction is correct,
for a given $\epsilon$, the BD correction fails for small enough $D$. We show in Figs. \ref{fig1} and \ref{fig2}
 a plot of $v_F$ and the  BD velocity  for $\epsilon=10^{-5}$, compared to the results of an exact
numerical calculation. In Fig. \ref{fig2} it can be seen as predicted that the BD treatment
does not apply for small $D$.  A calculation in this limit is presented in the next subsection.

\subsection{Small $D$, small $\epsilon$ limit}
\begin{figure}
\includegraphics[width=.4\textwidth]{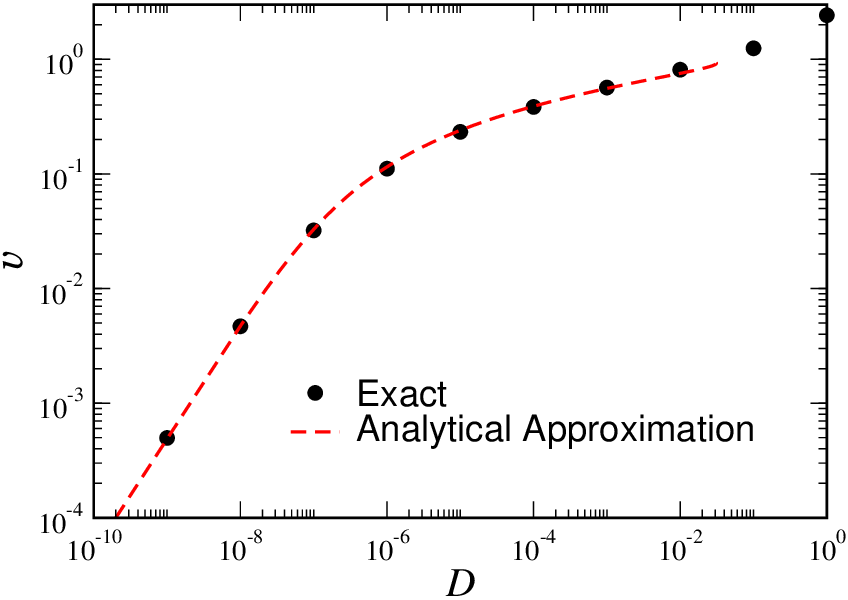}
\caption{Comparison of our analytic approximation, Eq. (\protect{\ref{result}}) and an exact numerical solution of
Eq. (\protect{\ref{basiceq}}). Parameters are $\epsilon=10^{-6}$, $\beta=r=1$. (color online)}
\label{fig3}
\end{figure}

Clearly, in the presence of a cutoff, the velocity should vanish as $D \to 0$.  Let us solve the model in this limit.  First, let us examine what happens when $D=0$.  Then, for small
$\epsilon$ we can
linearize around the solution $\phi_0=\frac{1}{1 + e^{rx/v}}$.  The equation reads
\begin{equation}
\left(\beta^2 - \frac{d^2}{dx^2}\right)\left(v\delta' + r\delta(1-2\phi_0)\right) = 0
\label{lineqn}
\end{equation}
This is equivalent to
\begin{equation}
v\delta' + r\delta\tanh\left(\frac{rx}{2v}\right) = Ae^{-\beta x} + Be^{\beta x}
\end{equation}
so that
\begin{equation}
\cosh^{-2}(\frac{rx}{2v})\left(v\delta \cosh^2\left(\frac{rx}{2v}\right)\right)' = Ae^{-\beta x} + Be^{\beta x}
\end{equation}
and
\begin{eqnarray}
\delta&=&\frac{1}{v \cosh^2\left(\frac{rx}{2v}\right)} \int_0^x \cosh^2\left(\frac{rx}{2v}\right) \left(Ae^{-\beta x} + Be^{\beta x}\right)
\nonumber\\
&=& \frac{1}{2v\cosh^2\left(\frac{rx}{2v}\right)}\left[ A\left( e^{-\beta x}\frac{\frac{r}{v}\sinh\left(\frac{rx}{v}\right) + 
\beta\cosh\left(\frac{rx}{v}\right)}{\left(\frac{r}{v}\right)^2-\beta^2} - \frac{\beta}{\left(\frac{r}{v}\right)^2-\beta^2}
+ \frac{1-e^{-\beta x}}{\beta}  \right)\right. \nonumber \\
&\ &\quad \quad\quad\quad\quad\quad\left.+ B\left( e^{\beta x}\frac{\frac{r}{v}\sinh\left(\frac{rx}{v}\right) -
\beta\cosh\left(\frac{rx}{v}\right)}{\left(\frac{r}{v}\right)^2-\beta^2} + \frac{\beta}{\left(\frac{r}{v}\right)^2-\beta^2}
+ \frac{e^{\beta x}-1}{\beta}  \right)\right]
\end{eqnarray}
We have chosen the limits of integration so that $\delta(0)=0$, so that the center of the front does not
move.  What is important is the large-$x$ asymptotics of $\delta$:
\begin{eqnarray}
\delta&\sim &\frac{A}{v}\left[e^{-\beta x}\left(\frac{\frac{r}{v}+\beta}{\left(\frac{r}{v}\right)^2-\beta^2}\right)
+2e^{-rx/v}\left(\frac{1}{\beta}-\frac{\beta}{\left(\frac{r}{v}\right)^2-\beta^2}\right)\right]\nonumber\\
&+& \frac{B}{v}\left[e^{\beta x}\left(\frac{\frac{r}{v}-\beta}{\left(\frac{r}{v}\right)^2-\beta^2}\right)
+2e^{-rx/v}\left(-\frac{1}{\beta}+\frac{\beta}{\left(\frac{r}{v}\right)^2-\beta^2}\right)\right]
\end{eqnarray}
We can now use the jump conditions, with $k_r=\beta$ since $D=0$, to fix $x_c$, $A$ and $B$.  We get, to leading order in $\epsilon$.
\begin{eqnarray}
e^{-rx_c/v}&=&\epsilon\frac{r}{r-\beta v} \nonumber\\
Ae^{-\beta x_c} &=& -\epsilon v \beta \nonumber \\
B &=& 0
\end{eqnarray}
The interesting question is now the behavior at $x\to -\infty$.  The leading asymptotics is
\begin{equation}
\delta\sim \frac{A}{v}e^{-\beta x}\left(\frac{-\frac{r}{v}+\beta}{\left(\frac{r}{v}\right)^2-\beta^2}\right)
\end{equation}
which of course violates the boundary conditions.  Thus, there is no solution without $D$.  To leading order in 
$D$, we get an inhomogeneous term, $D\beta^2 \phi_0''$, on the left-hand-side of Eq. (\ref{lineqn}). The inhomogeneous
solution, $\delta_D$, then satisfies the equation
\begin{equation}
v\delta_D' + r\tanh\left(\frac{rx}{2v} \right)\delta_D = \int_{-\infty}^\infty dy G(x-y) \left(\frac{-D\beta^2 r^2}{4 v^2}\right)\frac{\sinh\left(\frac{ry}{2v}\right)}{\cosh^3\left(\frac{ry}{2v}\right)}
\end{equation}
where $G$ is the Green's function for the operator $\beta^2 - \frac{d^2}{dx^2}$,
\begin{equation}
G(x-y)=\frac{1}{2\beta}e^{-\beta|x-y|}
\end{equation}
so that
\begin{equation}
\delta_D = \frac{1}{v\cosh^2\left(\frac{rx}{2v}\right)} \int_0^x dx' \cosh^2\left(\frac{rx'}{2v}\right) \int_{-\infty}^\infty dy G(x'-y)\left( \frac{-D\beta^2 r^2}{4 v^2}\right)\frac{\sinh\left(\frac{ry}{2v}\right)}{\cosh^3\left(\frac{ry}{2v}\right)}
\end{equation}
We need the asymptotic behavior of $\delta_D$ for large $x$.  For $r/v \gg \beta$, the integral is dominated
by the region of $x'$ large, $y\approx 0$. Thus,
\begin{eqnarray}
\delta_D &\sim& \frac{4 e^{-rx/v}}{v} \int_{-\infty}^x dx' \frac{e^{rx'/v}}{4} \int_{-\infty}^\infty dy
\frac{e^{-\beta x'}}{2\beta} \left(1 + \beta y + \frac{\beta^2 y^2}{2} + \ldots\right)\left( \frac{-D\beta^2 r^2}{4 v^2}\right)\frac{\sinh\left(\frac{ry}{2v}\right)}{\cosh^3\left(\frac{ry}{2v}\right)}\nonumber\\
&=&-\frac{D\beta r^2}{8v^3}e^{-rx/v}\int_{-\infty}^x dx' e^{rx'/v}e^{-\beta x'} \left( \beta\left(\frac{2v}{r}\right)^2 + \frac{\beta^3}{8}\left(\frac{2v}{r}\right)^4 \frac{\pi^2}{4}
+ \ldots\right) \nonumber \\
&=& -\frac{D\beta^2}{2v}\frac{1}{\frac{r}{v}-\beta}e^{-\beta x} \left(1 + \frac{\beta^2 v^2 \pi^2}{8r^2} + \ldots\right) 
\end{eqnarray}
We now have to again solve the jump conditions with this new contribution.  The coefficient $A$ above is now
modified and includes a term which, up to linear order in $\beta v/r$, reads
\begin{equation}
A_D = \frac{\beta^2 D}{2}\left(1 + \frac{\beta v}{r}\right)
\end{equation}
The condition for a solution is that this cancels the $A$ we found above, so that 
\begin{equation}
\frac{\beta^2 D}{2}\left(1 + \frac{\beta v}{r}\right) = \epsilon v \beta e^{\beta x_c}
\end{equation} 
or
\begin{equation}
D = \frac{2\epsilon v}{\beta\left(1+\frac{\beta v}{r}\right)} e^{-\beta v/r \ln(\epsilon r/(r-\beta v))} =
\frac{2vr}{\beta(r+\beta v)}\epsilon^{1-\beta v/r} \left(\frac{ r}{r-\beta v}\right)^{-\beta v/r}
\label{result}
\end{equation}
Thus, for very small $D$, the velocity is equal to $D\beta/(2\epsilon)$, which is reminiscent of the
behavior of evolution models for very small mutation rates~\cite{evol1}.  The comparison between our analytic 
approximation and an exact numerical solution is shown in Fig. (\ref{fig3}).

\section{\label{snyder_sec} The Snyder Discrete-Time Model}
Recently, Snyder~\cite{sny} introduced a model of colony spreading which, in one variant, involved an exponentially-distributed
hopping similar to the model defined above.  The essential difference between her model and ours is that hers was a
discrete-time model.  In each time step, all the offspring performed a hop and the parental generation was removed.  The
number of offspring at a given site was given by a local logistic growth law, similar to that incorporated in the Fisher model.
Snyder performed numerical simulations and measured the velocity of propagation, both for the stochastic model, and for
the corresponding (uncutoff) reaction-diffusion system, and found a difference between these two velocities.
Due to its close correspondence to the present model under investigation, it is useful to derive analytically
the uncutoff velocity and the BD approximation to the cutoff velocity,
so as to make clear the mapping between the Snyder model and ours.

As always, to derive the uncutoff "Fisher" (marginally-stable) velocity, it is enough to consider
the linearized version of the Snyder model, which reads
\begin{equation}
\phi^{t+1}(x)=\frac{r\beta}{2} \int_{-\infty}^{\infty} \phi^t(y) e^{-\beta|y-x|} dy
\label{sny1}
\end{equation}
where $r_S$ is the average number of offspring per individual and $\phi^t(y)$ is the number of
individuals at site $y$ at (integer) time $t$.
In Fourier space our equation reads:
\begin{equation}
\phi^{t+1}(k)=\frac{r_S\beta^2}{\beta^2+k^2}\phi^t(k)
\end{equation}
which, starting from a $\delta$-function initial condition gives
\begin{equation}
\phi^{t}(k)=\left[\frac{r_S\beta^2}{\beta^2+k^2}\right]^t
\end{equation}
or, Fourier transforming back,
\begin{equation}
\phi^{t}(x)=\int_{-\infty}^{\infty} \left[\frac{r_S\beta^2}{\beta^2+k^2}\right]^t e^{ikx} \frac{dk}{2\pi}
\end{equation}
We want to calculate the velocity, so we are interested in $\phi(x=vt)$, where
we have to choose $v$ such that this is independent of $t$ for large $t$.
This gives us a saddle-point integral
\begin{equation}
\phi^{t}(vt)=\int_{-\infty}^{\infty} e^{t\ln\left(\frac{r_S\beta^2}{\beta^2+k^2}\right)} e^{ikx} \frac{dk}{2\pi}
\end{equation}
The saddle point is at $k_*$ where
\begin{equation}
v=-\frac{2ik_*}{\beta^2+k_*^2}
\end{equation}
The dominant contribution to the integral is given by evaluating the integrand
 at the saddle, giving
\begin{equation}
\exp\left(t\left(\ln\left(\frac{r_S\beta^2}{\beta^2 + k_*^2}\right) + ik_*v\right)\right)
\end{equation}
If this is to be independent of $t$, the term in the exponential must vanish:
\begin{equation}
\ln\left(\frac{r_S\beta^2}{\beta^2 + k_*^2}\right) + \frac{2k_*^2}{\beta^2 
+ k_*^2} = 0
\end{equation}
Clearly, $k$ is pure imaginary and proportional to $\beta$, so we write
\begin{equation}
k_* = i\sigma \beta
\end{equation}
so that $\sigma$ depends only on $r$ and satisfies
\begin{equation}
\ln\left(\frac{r}{1 - \sigma^2}\right) = \frac{2\sigma^2}{1 
- \sigma^2} = 0
\label{v1_sny_SP}
\end{equation}
and the velocity is
\begin{equation}
v_F=\frac{2\sigma}{\beta(1-\sigma^2)}
\label{v2_sny_SP}
\end{equation}
It is reassuring that this formula reproduces the velocity measured by Snyder for the one set of parameters
presented in her paper.
For $r_S$ near 1, $v \sim 2\sqrt{(r-1)/2}/\beta$, while for large $r_S$,
$v\sim \ln(r_S)/\beta$.  Of course, on dimensional grounds this is reasonable,
since $v$ is a velocity per round, which has units of length, and $r_S$ is
dimensionless.  We see that $r_S$ near 1 corresponds to the Fisher limit, equivalent to the large $D\beta^2/r$ limit
of our model, since the growth rate of the population is $r_S-1$, so that small $r_S-1$ corresponds to a large value
of our dimensionless control parameter.  On the other hand, for large $r_S$, the models differ since the diffusion never goes away entirely in the everyone hops Snyder model. It is also interesting to note that in the Snyder model, the only effect
of $\beta$ is to set the velocity scale, as opposed to the more complicated role of $\beta$ in our model.

In fact, there is another way to solve equation (\ref{sny1}). We assume that the dependence
of $\phi$ in $t$ and $y$ is
\begin{equation}
\phi^t(y)=\phi(y-vt),
\label{phisny}
\end{equation}
and
\begin{equation}
\phi^t(y)=e^{-\alpha(y-tv)}.
\label{phisny_sol}
\end{equation}
putting (\ref{phisny_sol}) in (\ref{sny1}) yields:
\begin{equation}
e^{\alpha v}=\frac{r\beta^2}{\beta^2-\alpha^2}.
\label{v_sny1}
\end{equation}
Taking the derivative by $\alpha$ of (\ref{v_sny1}) (according to the marginal
stability criterion), and dividing
it by (\ref{v_sny1}) yields:
\begin{equation}
v_F=\frac{2\alpha_F}{\beta^2-\alpha_F^2},
\label{v_F_sny}
\end{equation}
Eqs. (\ref{v_sny1}) and (\ref{v_F_sny}) are seen to be equivalent to Eqs. (\ref{v1_sny_SP}) and (\ref{v2_sny_SP}) upon defining $\sigma\equiv \alpha_F/\beta$.

\begin{figure}  
\includegraphics[width=.4\textwidth]{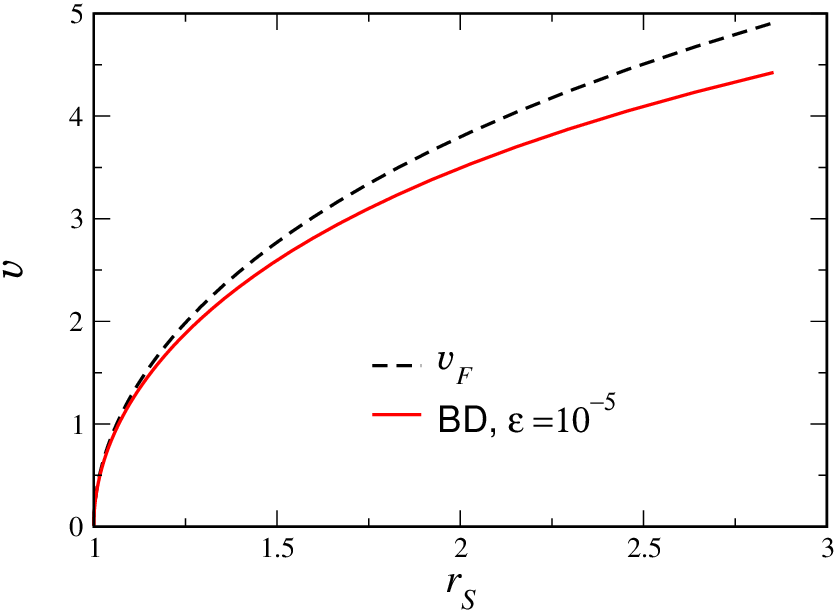} 
\caption{Snyder model: $v_F$ vs. $r_S$ according to  (\ref{v_F_sny}) and the BD correction
$v_{\epsilon}$ vs $r_S$. $\beta=0.5$ (color online)}
\label{fig4}
\end{figure}

We can eliminate $\alpha_F$ to obtain a direct relationship between $r$ and $v_F$ as follows:
\begin{equation}
\alpha_F=\frac{-1+\sqrt{v_F^2\beta^2+1}}{v_F}.
\label{alpha_F_sny}
\end{equation}
so
\begin{equation}
r=\frac{2(\sqrt{1+v_F^2\beta^2}-1)e^{\sqrt{v_F^2\beta^2+1}-1}}{v_F^2\beta^2},
\label{r_sny}
\end{equation}
The advantage of this second approach is we can immediately write down the BD correction, $v_{\epsilon}=v_F-\frac{v''(\alpha_F)\pi^2\alpha_F^2}{(ln\epsilon)^2}$. 
A graph of Snyder velocity with and without the correction is shown in Fig. \ref{fig4}.
We see that the larger $r_S$ is, the larger the correction
according to BD is, since as discussed above, increasing $r_S$ corresponds to decreasing the strength of diffusion in our model. 

\section{\label{grad_sec}Exponentially Distributed Hopping with a Reaction-Rate Gradient}

 In a previous work~\cite{kess,kess1},
 we studied the case of fronts 
propagating into an unstable state 
up a reaction-rate gradient.  
We focused again on the $A+B
\rightarrow 2A$ reaction~\cite{blumen}, with no $A$ particles and an
initial mean number $N$ of B particles at all sites past some initial
$x_0$, but with a reaction probability that depended linearly on spatial position.  This type of gradient would be a natural
consequence of spatial inhomogeneity, or could be imposed via a temperature 
gradient in a chemical reaction analog. 
Also, this type of system arises naturally in models of
Darwinian evolution~\cite{tsimring,rouzine}, (where 
fitness $x$ is the independent variable; the birth-rate, akin to
the reaction-rate here, is proportional to fitness). The naive equation describing such a 
model is the Fisher equation (\ref{fisher}) 
with a reaction strength $r=r_a(x)$ varying
linearly in space
\begin{equation}
r_a(x)  = \text{max}(r_{\text min},r_0+ \alpha x) \ .
\label{abs}
\end{equation} 
where $r_{\text min}$ is introduced to insure that the reaction rate stays
positive far behind the front, and has no effect on the velocity.
This model
gives rise to an accelerating front. We also introduced a {\em quasi-static} version of the model, 
wherein the reaction rate function moves along with the front:
\begin{equation}
r_q(x)  =  \text{max}(r_{\text min},\tilde{r}_0 + \alpha (x-x_f)) \ ,
\end{equation}
with $x_f$ is the instantaneous front position. This quasi-static problem should lead
 to a translation-invariant front with
fixed speed $v_q (\tilde{r}_0,\alpha )$.   Although important on its own, one might also try to view the
 quasi-static
 problem as a zeroth-order approximation to the original model, (the {\em absolute} gradient case),
  where by ignoring the acceleration, one obtains an adiabatic
approximation to the velocity $v (t; r_0, \alpha ) \simeq v_q
(\tilde{r} _0 (t), \alpha )$ with $\tilde{r} _0 (t) = r_0 +\alpha
x_f (t)$. In both models, fluctuations become
crucial due to the reaction gradient and the presence of the gradient
leads to a new class of fronts.  One characteristic of this new class is the
divergence of the front velocity with $N$. 
We found, that to leading order, 
the velocity of the front in the continuum limit 
diverges as $\ln^{1/3}(N)$, and to leading order on a lattice, the velocity diverges as $\sqrt{\ln{N}}$. It should be 
noted that in both
cases the leading order does not yield an accurate solution, and the next order correction must be taken into account.

Given that the nature of the divergence of the velocity with $N$ depends on the microscopic implementation
of diffusion (continuum versus lattice), it is natural to investigate this question for our model with exponentially
distributed hopping. Here, we chose to work on a lattice (with spacing $a$); we will see in the end that the results here are
not sensitive to the presence of the lattice. 
The model we study is:
\begin{eqnarray}
\frac{\partial \phi_i}{\partial t}&=&D \frac {(e^{\gamma}-1)^3}
{a^2 e^{\gamma}(e^{\gamma}+1)}
\left[\sum _{j=1}^{\infty}(e^{-\gamma j}(\phi_{i+j}+\phi_{i-j}))
-2\frac{\phi_i}{e^{\gamma}-1}\right ]
\nonumber\\
 &\ &{}+ r(i)\phi_i(1-\phi_i)\theta(p_i-\epsilon),
\label{eq2}
\end{eqnarray}
where $\gamma\equiv=\beta a$ is the rate of exponential falloff of the
hopping between successive lattice sites.
  It is easy to verify that this model reproduces continuum diffusion with coefficient $D$ for
 sufficiently smooth fields $\phi_i$.  We choose to focus on the quasi-static problem, as the
 presence of a steady-state solution makes the problem analytically tractable.  The steady-state solution
 on the lattice has the Slepyan~\cite{slepyan} form
 \begin{equation}
 \phi_i(t)=\phi(t - ia/v)
 \end{equation}
 so that each lattice point experiences the same history, with a time shift.  We define the
continuous variable $z=-v(t-ia/v)$, in terms of which 
 \begin{eqnarray}
0&=&D \frac {\left(e^{\gamma}-1\right)^3}{a^2 e^{\gamma}\left(e^{\gamma}+1\right)}
 \left[\sum_{j=1}^{\infty}\left[e^{-\gamma j}(\phi(z+aj)+\phi(z-aj))\right]
-2\frac{\phi(z)}{e^{\gamma}-1}\right]
\nonumber\\ &+&r(z)\phi(z)(1-\phi(z))\theta(p_z-\epsilon)+v \phi'(z),
\label{eq3}
\end{eqnarray}
 
  We wish to solve this equation for small $\epsilon$, assuming that $v$ will be large in this limit.
  Relying on our previous analysis of the nearest-neighbor hopping problem, we expect that the
  leading order solution for the velocity comes from the region of the front where $\phi$ is small, so the
  nonlinear $\phi^2$ term can be dropped. We assume~\cite{bender,rouzine,kess} a WKB-type solution
   $p_z=e^{S_z}$, and expand $S_{z \pm aj}$ into a Taylor series, so equation 
   (\ref{eq3})
  becomes:
 \begin{eqnarray}
0&=&D \frac {(e^{\gamma}-1)^3}{a^2 e^{\gamma}(e^{\gamma}+1)}\nonumber\\
&\times& \left[\frac{1}{e^{\gamma-S'a}-1}
 +\frac{1}{e^{\gamma+S'a}-1}-\frac{2}{e^{\gamma}-1}\right]
\nonumber\\ &+& r_0 +\alpha z+vS'.
\label{eq4}
\end{eqnarray}
After some algebra we get from (\ref{eq4}):
\begin{equation}
0=\frac{4D}{a^2}(e^{\gamma}-1)^2 \frac{\sinh^2(a S'/2)}{e^{2\gamma}+1
-2e^{\gamma}\cosh(S'a)}
+r_0 +\alpha z+vS'.
\label{eq5}
\end{equation}
As in the nearest-neighbor hopping problem, the only way to match to the post-cutoff solution is to
require that the front be close to the classical turning point.  In order to find the turning point, we need to equate the derivative of 
(\ref{eq4}) with respect to $S'$ to zero.
Doing so we get:
\begin{equation}
0=\frac{2D}{a}(e^{\gamma}-1)^4 \frac{\sinh(S'_*a)}{(e^{2\gamma}+1
-2e^{\gamma}\cosh(S'_*a))^2}
+v.
\label{stagstar}
\end{equation}
where $S'_*$ is the value of $S'$ at the turning point.
For large $\gamma$, Eq. (\ref{stagstar}) indeed matches the nearest-neighbor hopping result.
For large $v$, 
the denominator of the first term in Eq. (\ref{stagstar}) has to be close to $0$ in order
to balance the second term.
As the denominator vanishes if $aS'_*=-\gamma$, this gives
\begin{equation}
S'_*=-\frac{\gamma}{a} + \frac{\sqrt{\frac{D}{2a^3}e^{-2\gamma}\frac{(e^{\gamma}-1)^4}{\sinh(\gamma)}}}
{\sqrt{v}}
\label{epsi} 
\end{equation}
which is correct for $v\gg D\gamma/a$.
From (\ref{epsi}) one can obtain that, for fixed $\beta
\equiv \gamma/a$, $S'_*$ is only weakly dependent on $a$ for $0\le a\le 1$ as long as $\beta$ is
not too large, $\beta \lesssim 2$.
For example, for $\beta=1$, $S'_*=-1 + \sqrt{D/v}$ for $a \to 0$ and $-1 + 1.0019\sqrt{D/v}$ for $a=1$.  This
is reasonable, since the long-range nature of the hopping (for not too large $\beta$'s), smooths over
the lattice structure.

The fact that $|S'_*|$ is bounded by $\beta$ is the unique feature of our exponentially-distributed hopping.
We remind the reader that for standard continuum diffusion, $S'_*=-v/(2D)$, and so $|S'_*|$ grows unboundedly
with $v$, while for nearest-neighbor hopping, $|S'_*|$, though not linearly dependent,  still grows logarithmically with $v$.  
The faster the growth of $|S'_*|$,
the weaker the dependence of the velocity on $\ln(\epsilon)$.  This confirms our initial intuition that the exponentially distributed
hopping model should be more sensitive to fluctuations that even the nearest-neighbor hopping model.
It also reiterates why the lattice parameter $a$ is not important (for $\beta$ not large), since the rate of exponential
falloff of $\phi$ is bounded by $\beta$, and so never gets too large as to be affected by the lattice.

\begin{figure}  
\includegraphics[width=.4\textwidth]{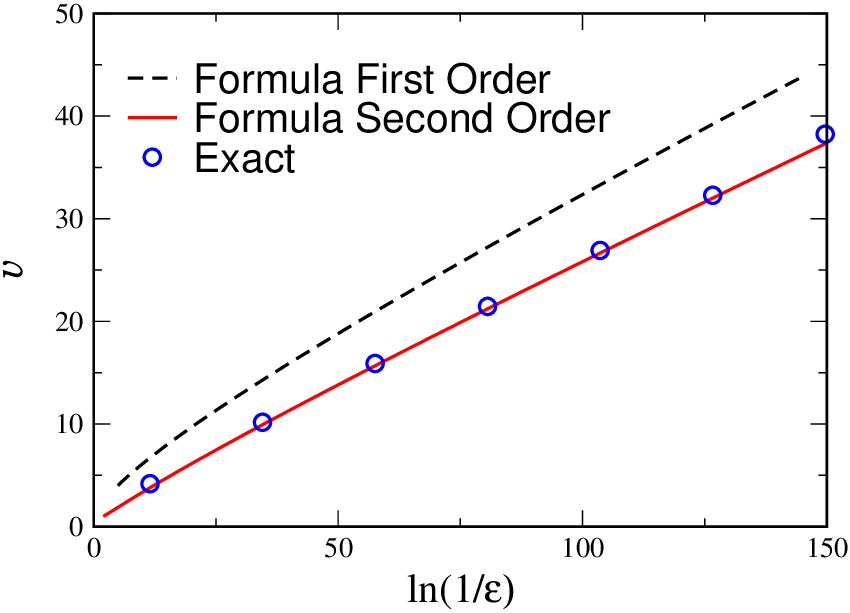} 
\caption{ $v$ vs. $ln(N)$ for $ D=1$, $\alpha =1$, $a=1$, $\alpha=0.1$, $B=1$. Numerical simulations
are compared to the 1st order formula, Eq. (\protect{\ref{sol}}) and the 2nd order formula, Eq. (\protect{\ref{int}}). (color online)} 
\label{fig6}
\end{figure}

Since the turning point is close to the cutoff point, the dominant contribution
 to the value of $\phi$ is $e^{S_*}$, where $S_*$ is the value of $S$ at the
turning point.
We now want to find $S_*$. This is given as :
\begin{eqnarray}
S_*&=& \int_0^{z_*}dz S'=\int_0^{S'_*}dS' S' \frac{dz}{dS'} \nonumber \\ 
&=&\int_0^{S'_*}dS'
S'(-\frac{1}{\alpha})\left[\frac{2D}{a}\frac{(e^{\gamma}-1)^4\sinh(\gamma)}
{(e^{2\gamma}+1-2e^{\gamma}\cosh(S'a))^2}+v\right]
\nonumber \\ 
&=& -\frac{D}{\alpha a^2e^{\gamma}}(e^{\gamma}-1)^4
\left[\frac{S'_*}{e^{2\gamma}+1-2e^{\gamma}\cosh(S'_*a)}\right]
\nonumber\\ 
&\ &{} - \frac{D}{\alpha a^2e^{\gamma}}(e^{\gamma}-1)^4\left[\frac{1}{a(e^{2\gamma}-1)}
 \ln\left(\frac{e^{\gamma+aS'_*}-1}{e^{\gamma}-e^{aS'_*}}\right)\right]
 \nonumber\\ 
 &\ &{} -\frac{1}{2\alpha}(S'_*)^2v
\label{int}
\end{eqnarray} 
To leading order, $\phi_c\equiv \phi(z_c)=e^{S_*}$.
In order to get the correction for $S_*$, we write, in the vicinity of the 
turning point,
\begin{equation}
\phi(z)=e^{S'_*z}\psi(z).
\label{corr} 
\end{equation}
Equation (\ref{corr}) smooths the variation between lattice points in the vicinity of
 the turning point, 
so we can expand $\psi(z)$ in a
Taylor series. This gives
\begin{eqnarray}
0&=&D \frac{(e^{\gamma}-1)^3}{a^2 e^{\gamma}(e^{\gamma}+1)}
\left[\sum _{j=1}^{\infty}(e^{-\gamma
j}(e^{S'_*aj}\psi(z+aj)+e^{-S'_*aj}\psi(z-aj)))\right]
 - 2D \frac{(e^{\gamma}-1)^3}{a^2 e^{\gamma}(e^{\gamma}+1)}\left[\frac{\psi(z)}
{e^{\gamma}-1}\right]
\nonumber\\
&\ &{} + (r_0 +\alpha z)\psi(z)+v\psi(z)S'+v\psi'(z)   \nonumber \\ 
&=&D \psi(z)\frac {(e^{\gamma}-1)^3}{a^2
               e^{\gamma}(e^{\gamma}+1)} \left[\frac{1}{e^{\gamma-S'_*}-1}\right]         \nonumber \\ 
&\ &{} + D \psi(z)\frac {(e^{\gamma}-1)^3}{a^2
          e^{\gamma}(e^{\gamma}+1)}\left[\frac{1}{e^{\gamma+S'_*}-1}-\frac{2}{e^{\gamma}-1}\right]  \nonumber\\ 
&\ &{} +  D \psi'(z)\frac {(e^{\gamma}-1)^3}{a^2
e^{\gamma}(e^{\gamma}+1)}\left[\left(\sum_j a j e^{-\gamma
j}(e^{a j S'_*}-e^{-a j S'_*})\right)\right] \nonumber\\ 
&\ &{} + D \psi''(z)\frac {(e^{\gamma}-1)^3}{a^2
e^{\gamma}(e^{\gamma}+1)}\left[\frac{1}{2}\left(\sum_j a^2 j^2 e^{-\gamma j}
(e^{ajS'_*}+e^{-ajS'_*})\right)\right]  \nonumber\\
&\ &{} +(r_0 +\alpha z)\psi(z)+v\psi(z)S'+v\psi'(z). 
\label{eqcorr}
\end{eqnarray}
 After some algebra, we get
\begin{eqnarray}
0&=&D \frac {(e^{\gamma}-1)^4\left[(e^{2\gamma}+1)\cosh(S'_* a)+2e^{\gamma}\cosh^2(S'_* a)
-4e^{\gamma}\right]}{
(e^{2\gamma}+1-2e^{\gamma}\cosh(S'_*a))^3}\psi''(z)
\nonumber\\
 &\ & {}+\alpha(z-z_*)\psi(z)
\label{eqcorr1} 
\end{eqnarray}
This is the Airy equation. The solution of (\ref{eqcorr1}) is
\begin{equation}
\psi(z)=\textrm{Ai}\left((z_*-z)\left(\frac{\alpha(e^{2\gamma}+1-2e^{\gamma}\cosh(S'_*a))^3}
{D(e^{\gamma}-1)^4\left[(e^{2\gamma}+1)\cosh(S'_* a)+2e^{\gamma}\cosh^2(S'_* a)-4e^{\gamma}\right]}\right)^{\frac{1}{3}}\right). 
\end{equation}
This gives us 
the distance from the turning point to the zero of
$\phi$. The first zero of the Airy function is at $-2.338$, so that the
 distance, $\ell$, is
\begin{equation}
\ell = 2.338\left(\frac{\alpha(e^{2\gamma}+1-2e^{\gamma}\cosh(S'_*a))^3}
{D(e^{\gamma}-1)^4\left[(e^{2\gamma}+1)\cosh(S'_*
a)+2e^{\gamma}\cosh^2(S'_* a)-4e^{\gamma}\right]}\right)^{-\frac{1}{3}} \ .
\end{equation} 
This gives us an addition contribution to $S_*$ of $S'_*\ell$. Adding this to
Eq. (\ref{int}) yields:
\begin{eqnarray}
\ln\left(\frac{1}{\epsilon}\right) &=&\frac{D}{\alpha a^2e^{\gamma}} (e^{\gamma}-1)^4
\left[\frac{S'_*}{e^{2\gamma}+1-2e^{\gamma}\cosh(S'_*a)}\right]
\nonumber\\
&-&\frac{D}{\alpha a^2e^{\gamma}} (e^{\gamma}-1)^4 \left[\frac{1}{a(e^{2\gamma}-1)}
\ln\left(\frac{e^{\gamma+aS'_*}-1}{e^{\gamma}-e^{aS'_*}}\right)\right]
\nonumber\\
 &+&\frac{1}{2\alpha}(S'_*)^2v
-2.338S'_*\left(\frac{\alpha}{D(e^{\gamma}-1)^4}\right)^{-\frac{1}{3}}\nonumber\\ 
&\times&  \left(\frac{(e^{2\gamma}+1-2e^{\gamma}\cosh(S'_*a))^3}{\left[(e^{2\gamma}+1)\cosh(S'_*
a)+2e^{\gamma}\cosh^2(S'_* a)-4e^{\gamma}\right]}\right)^{-\frac{1}{3}}
\label{sol}
\end{eqnarray} 
Again, this solution matches our previous solution for $\beta>>1$~\cite{kess,kess1}.
In the continuum limit, which as we noted above is accurate for $\beta \lesssim 2$,
this equation becomes
\begin{eqnarray}
\ln\left(\frac{1}{\epsilon}\right) &=&\frac{D\beta^4}{\alpha} 
\left[\frac{S'_*}{\beta^2-(S'_*)^2}\right]
 - \frac{D\beta^3}{2\alpha}\ln\left(\frac{\beta+S'_*}{\beta-S'_*}\right)
\nonumber\\
 &\ &{} + \frac{1}{2\alpha}(S'_*)^2v
-2.338S'_*\left(\frac{\alpha}{D\beta^4}\right)^{-\frac{1}{3}} \left(\frac{\left[\beta^2-(S'_*)^2\right]^3}{\beta^2 + 3(S'_*)^2}
\right)^{-\frac{1}{3}}
\label{sol-cont}
\end{eqnarray}
Substituting the continuum limit of our expression for $S'_*$ in the above yields, and expanding for large $v$ yields
\begin{eqnarray}
\ln\left(\frac{1}{\epsilon}\right) &=& \frac{\beta^2 v}{2\alpha} + \sqrt{v}\left[- \frac{\sqrt{2D\beta^5}}{\alpha} + 2.338\frac{2^{1/6}\sqrt{\beta}}{\alpha^{1/3}D^{1/6}}\right] + \frac{D\beta^3}{4\alpha}(2 + \ln\left(\frac{8 v}{D\beta}\right)) - 2.338\frac{2^{2/3}D^{1/3}\beta}{\alpha^{1/3}}
\end{eqnarray}
which is still fairly messy.  To test these formulas, we present in Fig. \ref{fig6} the velocity versus $\ln(1/\epsilon)$, comparing between Eqs. (\ref{sol}) and (\ref{int})
and numerical results from direct integration of the time-dependent equation.  We see that the agreement between theory and
simulation is quite good, and that the correction term is not negligible for this range of $\ln(1/\epsilon)$.

\begin{figure}  
\includegraphics[width=.4\textwidth]{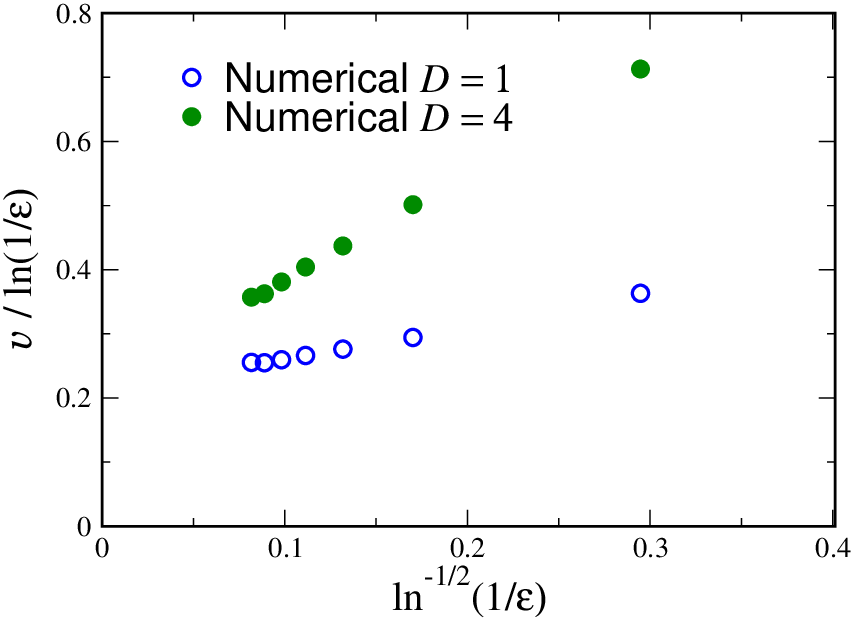} 
\caption{ $v/ln(N)$ vs. $ln(N)$ for $ D=1$, $\alpha =1$, $a=1$, $\alpha=0.1$, $r_0=1$. The data presented
are from numerical simulations of the cutoff deterministic equation. (color online)} 
\label{fig6a}
\end{figure}

The first interesting thing to note about our analytic result is that asymptotically, for small cutoff, the velocity is proportional to
$\ln(1/\epsilon)$, with a coefficient {\em independent} of $D$.  This is reminiscent of the "velocity without
diffusion" we saw in the zero-gradient case in the absence of a cutoff.  We can see this point clearly in Fig \ref{fig6a} where we
graph $v/\ln(1/\epsilon)$ as a function of $1/\sqrt{\ln(1/\epsilon)}$ for $D=1$ and $D=4$.  It is clear that the two graphs are
converging to the same value of $\beta^2/(2\alpha)=0.2$.

\begin{figure}  
\includegraphics[width=.4\textwidth]{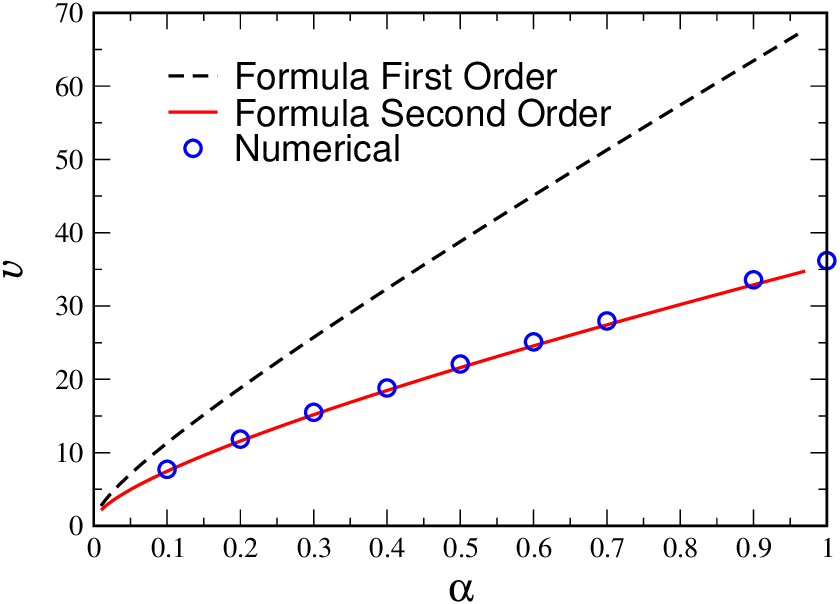} 
\caption{ $v$ vs. $\alpha$ for $ D=1$, $\beta =1$, $a=1$, $r_0=1$, $ln(N)=25$. Numerical simulations
are compared to the 1st order formula, Eq. (\protect{\ref{sol}}) and the 2nd order formula, Eq. (\protect{\ref{int}}).(color online)}
\label{fig7}
\end{figure}

In Fig. \ref{fig7} we present results for $v$ versus $\alpha$, again comparing the analytic formulas Eqs. 
(\ref{sol}) and (\ref{int}) to the results of direct simulation. Again the agreement is very satisfying.   For large $\alpha$
the "correction" term is dominant and the velocity grows as.      
\begin{equation}
v \sim 0.1452\frac{ln^2(1/\epsilon)\alpha^{2/3}D^{1/3}}{\beta}
\label{sol_large_a}
\end{equation}  
Unfortunately, this asymptotic result is only valid for extremely large $\alpha\gg \ln^3(1/\epsilon)$.   
                                                                                     
\begin{figure}  
\includegraphics[width=.4\textwidth]{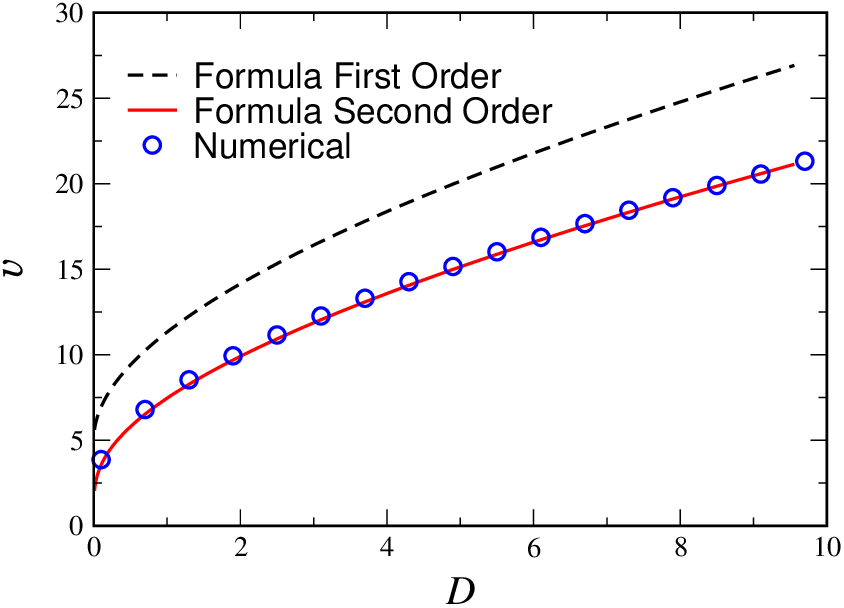} 
\caption{ $v$ vs. $D$ for  $\beta =1$, $a=1$, $\alpha=0.1$, $r_0=1$, $ln(1/\epsilon)=25$. Numerical simulations
are compared to the 1st order formula, Eq. (\protect{\ref{sol}}) and the 2nd order formula, Eq. (\protect{\ref{int}}).
(color online)}
\label{fig8}
\end{figure}

\begin{figure}  
\includegraphics[width=.4\textwidth]{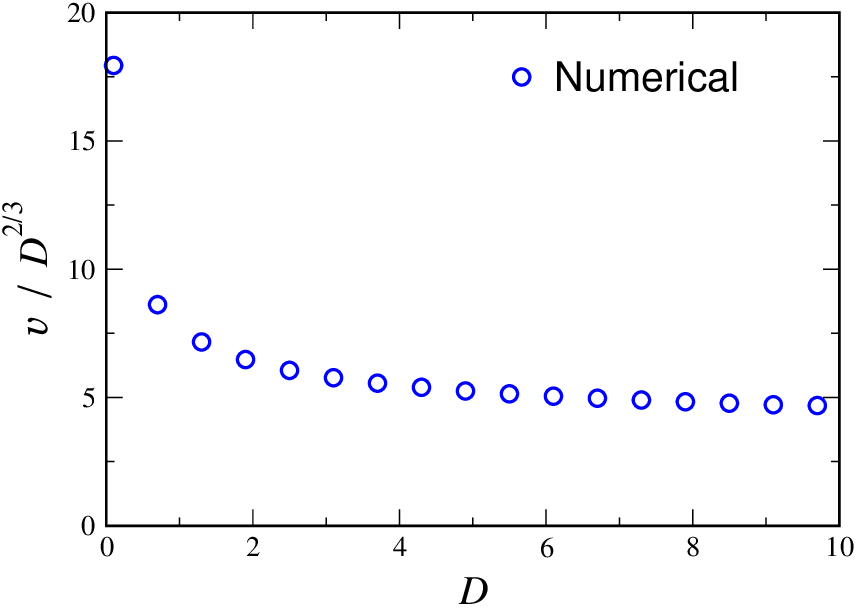} 
\caption{ $v/D^{2/3}$ vs. $D$ for  $\beta =1$, $a=1$, $\alpha=0.1$, $r_0=1$, $ln(1/\epsilon)=25$. Data
presented are from numerical simulations of the deterministic cutoff equation. (color online)}
\label{fig8a}
\end{figure}

The dependence of the velocity on the diffusion constant $D$ is presented in  Fig. \ref{fig8}, where we again
present a comparison with our theoretical prediction.  It is seen that as $D$ grows, $v/D$ decreases, and so
our analytic approximation for $S'_*$ becomes increasing less reliable.  Further analysis shows that in fact
for very large $D$, $S'_* \approx v/(2D)  \ll 1$, and the calculation reverts to that of the standard continuum
diffusion presented in Ref. \onlinecite{kess}, where $v \sim f(\alpha,\epsilon)D^{2/3}$, and the
prefactor $f \sim (24\alpha\ln(1/\epsilon))^{1/3}$ for $\epsilon\to 0$. We can verify this result
by replotting the data in Fig. \ref{fig8a}, this time showing $v/D^{2/3}$, which is seen to be consistent with an
approach to a constant close to $(24\alpha\ln(1\epsilon))^{1/3}=3.91$. This reversion to continuum diffusion for large $D$
is reasonable, 
since if diffusion is fast enough, it is irrelevant
how it is implemented.  For extremely small $D$ our calculation becomes unreliable, since there one is not allowed to truncate to an Airy equation.  We
expect, similar to what we occurs in the evolution problem, that the velocity will be proportional to $D/\epsilon$ in this limit.

\begin{figure}  
\includegraphics[width=.4\textwidth]{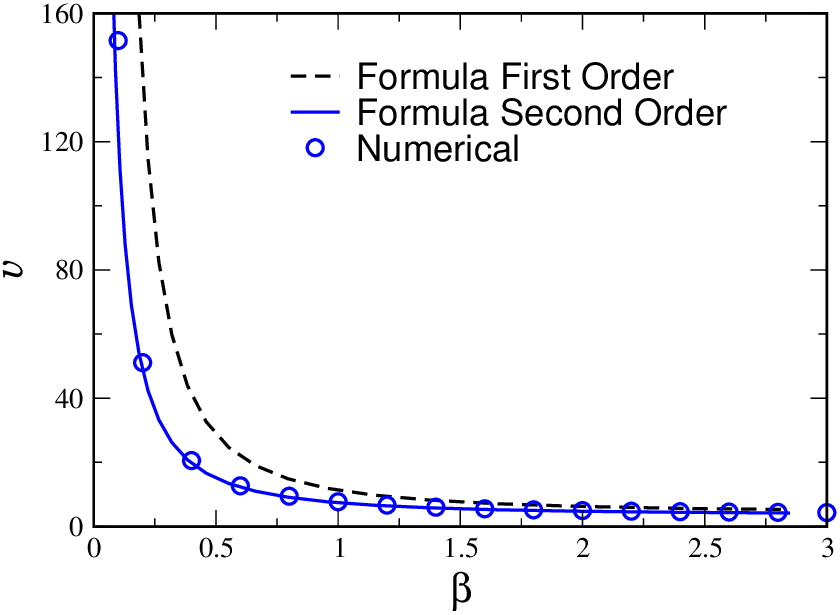} 
\caption{ $v$ vs. $\beta$ for $D=1$, $a=1$, $\alpha=0.1$, $r_0=1$, $ln(1/\epsilon)=25$. Numerical simulations
are compared to the 1st order formula, Eq. (\protect{\ref{sol}}) and the 2nd order formula, Eq. (\protect{\ref{int}}).
(color online)}
\label{fig9}
\end{figure}

\begin{figure}  
\includegraphics[width=.4\textwidth]{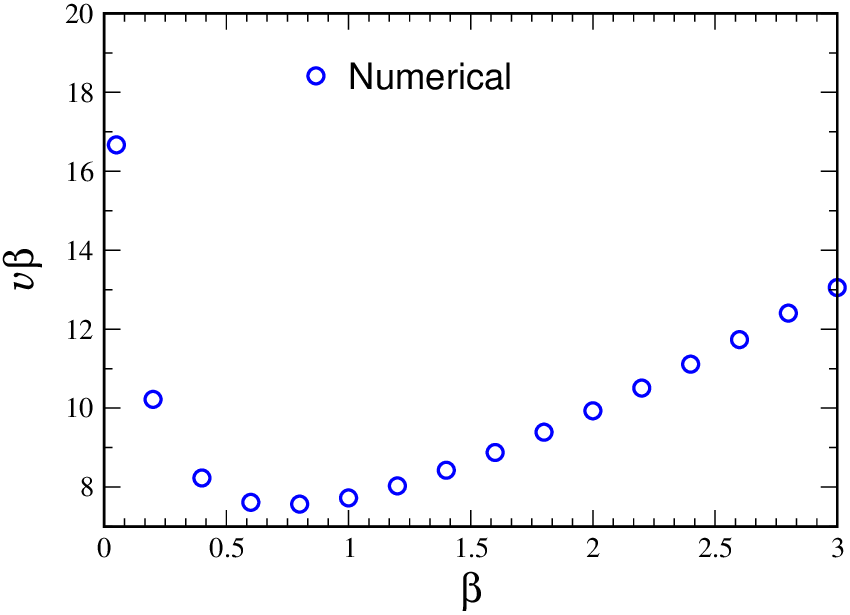} 
\caption{ $v\beta$ vs. $\beta$ for $D=1$, $a=1$. $\alpha=0.1$, $r_0=1$, $ln(1/\epsilon)=25$. Data
presented are from numerical simulations of the deterministic cutoff equations. (color online)}
\label{fig9a}
\end{figure}

Lastly, In Fig. \ref{fig9}, we can see a comparison between (\ref{sol}), (\ref{int})
and numerical results for $v$ vs. the rate of falloff of the hopping distribution, $\beta$.                                                     
For large $\beta$, the problem reverts to the nearest neighbor hopping model, so $v$
should approach a constant in that limit, consistent with the data presented.  For small $\gamma$,
again the "correction" term is dominant and we recover the large $\alpha$ result, Eq. (\ref{sol_large_a}),
with $v$ diverging as $1/\beta$.  We therefore plot $v\beta$ versus $\beta$ in Fig. \ref{fig9a},
where we see that the data is consistent with $v\beta$ approaching the constant 
$0.1452\ln^2(1/\epsilon)\alpha^{2/3}D^{1/3}=19.55$ for small $\beta$.

\begin{figure}  
\includegraphics[width=.4\textwidth]{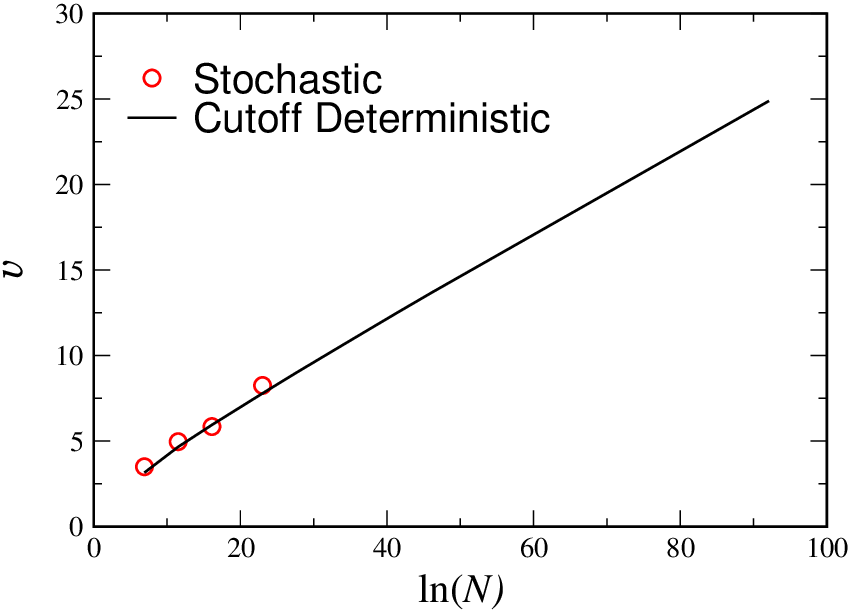} 
\caption{ Comparison between stochastic results 
and numerical results for $v$ vs. $ln(N)$ for $D=1$, $a=1$, $\alpha=0.1$, $r_0=1$.
$x_f=\sum{\phi_k}$. (color online)}
\label{fig11}
\end{figure}

The last task before us is to test our cutoff theory is a good approximation for
the stochastic case. The analytical
procedure done above is referring to the case for which the front
 position $x_f$ is defined to by $\phi(x_f)=1/2$.  For the stochastic case this procedure
 is ill-defined, since $\phi$ fluctuates.  Rather, we choose to define the front
by
\begin{equation}
x_f=\sum_k \phi_k \ , 
\end{equation}
Rather than redo the theory for this definition of the front, we chose the expedient
of comparing the
the stochastic results to
 numerical results that also define the 
front position as the sum of $\phi_k$, which amounts to a shift in $r_0$.
The comparison is shown in Fig. \ref{fig11}.

As a closing remark, we note that one of the most interesting aspects of the above calculation
(and the previously published calculations for the nearest neighbor hopping model) is that
the result does not at all depend on form of the solution past the cutoff point;
the mere existence of a cutoff is enough to force the system to the
WKB turning point and hence fix the velocity.

\section{Summary}
\label{final_sec}
We have investigated herein
reaction-diffusion systems in which the hopping probability
exponentially decays with distance, focussing on the fluctuation induced anomalies
seen in the same systems with continuous diffusion and nearest neighbor hopping.  As in
these previously studied cases, we probe the sensitivity to fluctuations by studying the
dependence of the steady-state velocity on a cutoff in the reaction term when the
density drops below a cutoff of the order of one particle per site. We first studied this model with no gradient,
 showing that, in the absence of a cutoff, the velocity does not vanish for small $D$.
 We showed that the BD correction for velocity due to the presence of a
 cutoff diverges in the case of
  small $D$, and calculate that the velocity actually vanishes linearly 
  for small $D$ in the presence of cutoff. 
 Our model is similar to a discrete-time model describing the
spread of colonies, and we show the same generic features apply to this model as well. 
We then studied the effect of introducing a quasi-static gradient into our model. Here, even for
continuum diffusion and nearest-neighbor hoppings, fluctuation effects lead to a divergence of
the velocity with increasing particle density $N$.  We found that this phenomenon is enhanced
by the exponential distributed hopping, so that the velocity diverges more strongly,
as $\ln(1/\epsilon)$.  In fact, for long-range hopping, $\beta \ll 1$, the velocity is proportional
to $\ln^2(1/\epsilon)$.
Our analytical work was confirmed by direct simulation of the cutoff deterministic equation,
as well as by comparison to the original stochastic model.

\acknowledgments{We acknowledge the support of the Israel Science Foundation.  We thank
Herbert Levine for useful discussions.}

\appendix*
\section{Numerical Simulations}

In the body of the paper, we have presented results from direct numerical simulations of both
the deterministic cutoff reaction-diffusion equation and the stochastic particle model.  Here
we briefly present some relevant details of the simulation methods, especially in reference to treating
in an efficient manner the long-range nature of the hopping.

\subsection{Deterministic Equation}
The simulations are essentially standard, using an Euler method time step. The only subtlety is
in handling the hopping term efficiently.  A naive treatment would involve calculating the
transfer of density from every pair of sites, which is a prohibitively expensive $O(L^2)$ operation, where $L$ is the
spatial extent of the lattice.

To solve this difficulty, consider the density transfered to site $i$ from all the sites to the left; i.e.,
$1,2\ldots i-1$. This transfered density, which we denote $L_i$ is given by
 \begin{equation}
 L_{i}=\sum_{j=1}^{i-1}\phi_{j} e^{-\gamma(i-j)}
 \end{equation}
 $L_i$ satisfies a simple recursion relation:
 \begin{equation} 
 L_i=(L_{i-1} + \phi_{i-1}) e^{-\gamma}
 \end{equation}
 Thus, in one pass we can calculate how much density is transferred to every site from all its left neighbors.
 The density transferred from the right neighbors is done similarly, using 
 \begin{equation}
 R_i=\sum_{j=i+1}^{L} \phi_j e^{-\gamma(j-i)}
 \end{equation}
 and the recursion relation
 \begin{equation}
 R_i=(R_{i+1} + \phi_{i+1}) e^{-\gamma}
 \end{equation}
 and making a leftward pass over the sites.
The problem is thus reduced to an $O(L)$ problem.

\subsection{Stochastic Simulation}

Our basic technique for simulating the stochastic model is to treat
all the particles on a given site in "bulk"~\cite{kns,kess}.  The number of
particles that participate in any given process (birth, death and hopping)
is given by a binomial distribution, and so can be determined by
drawing a binomial deviate.  The simulation performs in parallel first
  a  hopping step, followed
  by a reaction step.
   In the reaction step, the
  number of $B$ particles which transform into $A$'s at site $x$
  is again a binomial deviate, drawn from
  $B(N_B(x),1-\left(1-r(x)dt\right)^{N_A(x)})$.
  Replacing the distribution by its expected value, and setting $N_B(x)=
  N - N_A(x)$, and defining $\phi=N_A/N$
  gives Eq. (\ref{fisher}).  A $dt$ small
  enough so that less than 10\% of the $A$, $B$'s at a site hop and/or react in
  one time step is sufficient; smaller values do not alter the results.

Again, hopping in our model provides a challenge,
since we cannot afford to draw a binomial deviate for every pair of sites. Rather,  
every time step we first determine the
number of particles {\em{leaving}} that site due to the hopping, by drawing a single
binomial deviate. We then determine how many of these move to the right, by drawing
a second deviate. Of those moving to the left (right), we determine how many move to the
nearest neighbor, by drawing a third deviate, and remove this number from the pool of left (right)
movers.  Then, if any particles remain in the pool, we determine how many move to the second
nearest neighbor, removing these from the pool, continuing in this manner till the pool is exhausted.
The number of deviates we need to choose is thus fixed (on average) by $\gamma$, independent of $L$.

\end{document}